\documentclass[letterpaper]{IEEEtran}  
\usepackage[french,english]{babel}
\usepackage{amsmath,amsfonts,amssymb,bm,mathrsfs}
\usepackage{graphicx,color}
\usepackage[utf8]{inputenc}
\usepackage[T1]{fontenc}
\graphicspath{{./Figs/}}

\newcommand{\xp}[1]{\mathbb{E}\left\{#1\right\}}

\newcommand{\cor}[1]{\textcolor{black}{#1}}

\definecolor{MidGreen}{rgb}{0,0.65,0}
\definecolor{DarkGreen}{rgb}{0,0.40,0}
\definecolor{MidBlue}{rgb}{0,0,0.75}
\definecolor{DarkBlue}{rgb}{0,0,0.52}
\newcommand{\Enrico}[1]{\textcolor{black}{#1}}

\title{Response and Uncertainty of the Parabolic Variance PVAR to Non-Integer Exponents of the Power Law}

\author{François Vernotte, Siyuan Chen, Enrico Rubiola
\thanks{F. Vernotte and E. Rubiola are with (1) FEMTO-ST Institute, CNRS Lab no.~6174, Time and Frequency Department, (2) Observatory THETA, and (3) UBFC, all in Besançon, France.  E. Rubiola is also with the Istituto Nazionale di Ricerca Metrologica INRiM, Divsion of Quantum Metrology and Nanotechnology, Torino, Italy.
Siyuan Chen is with (1) Radioastronomy Station Nançay, PSL, Nançay, France, (2) LPC2E, Université d'Orléans, France, (3) FEMTO-ST Institute, Time and Frequency Department, (4) UBFC.}}

\begin{document}
\maketitle


\begin{abstract}\boldmath
Oscillator fluctuations are described as the phase or frequency noise spectrum, or in terms of a wavelet variance as a function of the measurement time.  The spectrum is generally approximated by the `power law,' i.e., a Laurent polynomial with integer exponents of the frequency.  
\Enrico{This article extends the domain of application of PVAR, a wavelet variance which uses the linear regression on phase data to estimate the frequency, and called `parabolic' because such regression is equivalent to a parabolic-shaped weight function applied to frequency fluctuations.
In turn, PVAR is relevant in that it improves on the widely-used Modified Allan variance (MVAR) enabling the detection of the same noise processes at the same confidence level in a shorter measurement time. 
More specifically, we provide (i) the analytical expression of the response of the PVAR to the frequency-noise spectrum in the general case of non-integer exponents of the frequency, and (ii) a useful approximate expression of the statistical uncertainty.}

\end{abstract}


\section{Introduction}
The fluctuations of an oscillator are generally described as the phase noise $\mathscr{L}(f)$\cor{, where $f$ is the Fourier frequency}, or as the two-sample variance $\sigma^2_\mathsf{y}(\tau)$\cor{, where $\tau$ is the integration time}. 
The latter takes different flavors, the most known of which are the Allan variance (AVAR) and the modified Allan variance (MVAR)\@.  
The concepts related to $\mathscr{L}(f)$ were introduced in the 1960s to describe the fast fluctuations of oscillators for radars and frequency synthesis \cite{Chi-1965}.  By contrast, $\sigma^2_\mathsf{y}(\tau)$ was introduced to describe the fluctuations of Cs-beam clocks for timekeeping, with obvious focus on slow fluctuations \cite{Cutler-1966,Allan-1966}. 
Traditionally, the boundary between these two choices was $\tau\approx0.1\ldots1$, or $f\approx1\ldots10$ Hz.  The overlap was rather small, of the order of one decade.  In fact, time counters could not be easily used at a sampling interval $\tau_0$ smaller than $\approx100$ ms, limited by the slowness of the IEEE 488 BUS transferring ASCII 
 data.  By contrast, the measurement of $\mathscr{L}(f)$ at low Fourier frequencies was limited by the narrow dynamic range of the double balanced mixer used as the phase-to-voltage converter (no more than $\pm20^\circ$), and of the \cor{analog to digital converters}.  The \cor{Fast Fourier Transform} analyzers were so complex and expensive that they were avoided when possible.  Interestingly, the two-sample variance is broadly equivalent to a one-octave filter centered at $f\approx0.45/\tau$. 

Nowadays these limitations are gone, and the overlap in the domain of application of $\mathscr{L}(f)$ and $\sigma^2_\mathsf{y}(\tau)$ is of 6--8 decades. 
Digital instruments can measure $\mathscr{L}(f)$ from 0.1--1 mHz \cite{TimePod,PhaseStation,FSWP,Feldhaus-2016}.
This is made possible by Software Defined Radio techniques (see \cite{SDR1,SDR2} for a general overview), which enable phase measurements not bounded to $\pm\pi$, \Enrico{and low sampling frequency by proper decimation of high-speed data}.  The CORDIC algorithm \cite{Volder-1959,Meher-2009} is the preferred choice to calculate $\varphi(t)$ from the digitized I/Q stream.
Counters with picosecond resolution were available since the 1970s with the Nutt interpolator \cite{Nutt-1968}, but continuous time stamps at a sampling interval $\tau_0\approx100$ ns \cite{Carmel,Guidetech} could be possible only thanks to \cor{Field Programmable Gate Arrays} (FPGAs). 
The minimum $\tau$ is actually greater than $\tau_0$ because trivial limitations intervene, but the practical limit is still of the order of several $\mu$s.
The conclusion is that assessing the \emph{equivalence between spectra and variances} is more important than ever.

It is generally agreed that the phase noise of oscillators is well described by the `power law’ or `polynomial law' model, which is the extension of the regular polynomial to the negative powers of the variable (Laurent polynomials). 
While the literature is shy about exceptions, we came across significant practical cases where the phase noise has a non-integer slope over a few decades.
In other domains of physics, \Enrico{the term `flicker noise' refers to a noise process whose spectrum is of the $f^\beta$ type}, where the exponent $\beta$ is actually in $[-1.2,-0.8]$ to $[-1.5,-0.5]$ depending on the author \cite{Dutta-1981,Weissman-1988,Milotti-1995}.  Accordingly, we may find $f^\beta$ phase noise \Enrico{in oscillators, and} $f^{\beta-2}$ phase noise after the phase-to-frequency conversion known as the Leeson effect \cite{Rubiola-2008-Leeson}.  The fractional-order frequency control, nowadays quite popular \cite{Chen-2009,Petras-2011,Sheard-2003}, is a good reason for non-integer slopes \Enrico{to be present} in the spectrum of a locked oscillator or laser.  Non-integer slopes also appear in other branches of frequency metrology.  For example, theoretical predictions about millisecond pulsars suggest that the common FM noise could follow the $f^{-7/3}$ law \cite{Phinney-2001,Chen-2017}. Finally, a continuous polynomial law is necessary in Bayesian statistical analysis, \Enrico{when we estimate the polynomial law from the measured spectrum} \cite{Chen-2020}.  \Enrico{Interestingly, the continuous law is needed as an intermediate step even when estimation targets integer exponents.
}

The response of $\sigma^2_\mathsf{y}(\tau)$ to phase noise in the case of non-integer exponents of the power law was already solved for \cor{the Allan Variance (AVAR)} and \cor{the Modified Allan Variance (MVAR)} \cite{Walter-1994}, while \cor{the Parabolic Variance (PVAR)} was introduced later \cite{Benkler-2015,Vernotte-2016-TUFFC}. 
\Enrico{In our opinion, MVAR is obsoleted by PVAR} because PVAR is suitable to the same applications, \Enrico{and it enables} the detection of the same noise phenomena\Enrico{, in the same conditions} at the same confidence level with a smaller data record \cite{Vernotte-2016-TUFFC}, \Enrico{i.e., in a shorter measurement time}.

This work stands on \cite{Walter-1994} and extends the results to PVAR providing conversion formulae, degrees of freedom and statistical uncertainty (Type A uncertainty, according to the definitions given by the \cor{International Vocabulary of Metrology} `VIM' \cite{VIM}).

\section{The Response to Polynomial Spectra}
\subsection{Basic Definitions and Tools}
We consider a clock signal $V_0\cos[\omega_0t+\varphi(t)]$ of nominal frequency $\omega_0/2\pi$ and random phase $\varphi(t)$.  It is understood that $\varphi(t)$ is not bound to $\pm\pi$, and that $|\dot{\varphi}(t)|\lll\omega_0$.   The associated time fluctuation $\mathsf{x}(t)=\varphi(t)/\omega_0$ is usually referred to as \emph{phase time}.  The quantity $\mathsf{y}(t)=\dot{\mathsf{x}}(t)$ is the fractional frequency fluctuation.

According to the IEEE Standard 1139 \cite{IEEE-STD-1139-2008}, the phase noise is defined as $\mathscr{L}(f)=\frac{1}{2}S_\varphi(f)$, that is, half of the single-sided Power Spectral Density (PSD) of $\varphi(t)$.  For our purposes, \cor{it is convenient to use the quantity}
\begin{equation}
  S_\mathsf{y}(f)=\frac{f^2}{(\omega_0/2\pi)^2}S_\varphi(f)~,
\end{equation}
\cor{instead of $S_\varphi(f)$, which provides fully equivalent information.}  The associated polynomial law is usually written as
\begin{equation}
  S_\mathsf{y}(f)=\sum_{\alpha=-2}^{2}\mathsf{h}_{\alpha} f^\alpha~,
  \label{eqn:poly-law}
\end{equation}
where the exponent $\alpha$ equals $-2$ for random walk FM noise, $-1$ for flicker FM noise, $0$ for white FM noise, $1$ for flicker PM noise, and $2$ for white PM noise.  

From a general perspective, the two-sample variance can be written as 
\begin{equation}
  \sigma^2_\mathsf{y}(\tau)=
  \frac{1}{2}\mathbb{E}\Bigl\{\bigl[\overline{\mathsf{y}}_2-\overline{\mathsf{y}}_1\bigr]^2\Bigr\}~,
  \label{eqn:2s-var}
\end{equation}
where $\mathbb{E}\{\:\}$ is the mathematical expectation, \cor{ and
$\overline{\mathsf{y}}_1$ and $\overline{\mathsf{y}}_2$ are the two samples of $\mathsf{y}(t)$ } averaged over contiguous time intervals of duration $\tau$ (hereafter the integration time).  
Our use of \eqref{eqn:2s-var} differs from the general literature in that $\overline{\mathsf{y}}_2$ and $\overline{\mathsf{y}}_1$ are weighted averages. The uniform average gives AVAR, the triangular\Enrico{-weight} average gives MVAR, and \Enrico{the parabolic-weight average gives PVAR.} 
Other options are possible, for example the Hadamard and the Picimbono variances.  Accordingly, \eqref{eqn:2s-var} is rewritten as
\begin{equation}
  \sigma^2_\mathsf{y}(\tau) = 
  \mathbb{E} \Bigl\{ \Big[
  \int_{-\infty}^{\infty} \mathsf{y}(t)\:w(t)\:\mathrm{d}t
  \Bigr]^2\Bigr\}~,
\label{eqn:wvar}
\end{equation}
where $w(t)$ is a wavelet-like function that describes $\overline{\mathsf{y}}_2-\overline{\mathsf{y}}_1$, including the weight functions.
The specific $w(t)$, named $w_A(t)$ for AVAR, $w_M(t)$ for MVAR and $w_P(t)$ for PVAR are defined in \cite[Fig.~3 and related text]{Vernotte-2016-TUFFC}. \cor{For example, the PVAR weight function} 
\cor{is 
\begin{equation}
w_P(t)=\frac{3\sqrt{2}\,t}{\tau^{3}}\left(|t|-\tau\right)\label{eq:wght}
\end{equation}
with $t\in[-\tau,\tau]$.} 
\Enrico{This is a parabola, which we refer to as $\Omega$, the most similar Greek letter.} 
\cor{Since $\mathsf{y}(t)$ is the derivative of $\mathsf{x}(t)$, Eq.~\eqref{eqn:wvar} can be rewritten as
\begin{equation}
  \sigma^2_\mathsf{y}(\tau) = 
  \mathbb{E} \Bigl\{ \Big[
  \int_{-\infty}^{\infty} \mathsf{x}(t)\:\dot{w}(t)\:\mathrm{d}t
  \Bigr]^2\Bigr\}~,
\label{eqn:wvarx}
\end{equation}}
\cor{where $\dot{w}(t)$ is the time derivative of $w(t)$. Thus,} \Enrico{it holds that}
\cor{%
\begin{equation}
\dot{w}_P(t)=\frac{6\sqrt{2}}{\tau^{3}}\left(|t|-\frac{\tau}{2}\right)\label{eq:wghtx}
\end{equation}}%
\Enrico{for PVAR}, with $t\in[-\tau,\tau]$.

\cor{In practice, the variance is calculated from a stream of $N$ samples $\mathsf{x}_j$ \Enrico{regularly spaced by $\tau_0$, which gives the measurement time $\tau=m\tau_0$, integer $m$} (hereafter, the normalized integration time).}
The expectation  $\mathbb{E}$ is replaced with the average $\left<\;\right>_M$ of $M$ realizations of $\overline{\mathsf{y}}_2-\overline{\mathsf{y}}_1$, and $\sigma^2_\mathsf{y}(\tau)$ is replaced with $\mathrm{AVAR}(\tau)$ or $\mathrm{PVAR}(\tau)$
\begin{align}
  \mathrm{AVAR}(\tau)&=
  \frac{1}{2M}\sum_{i=0}^{M-1} \Bigl[\overline{\mathsf{y}}_{i+1}-\overline{\mathsf{y}}_{i}\Bigr]^2\\[1ex]
  \mathrm{PVAR}(\tau)
  &=\frac{72}{Mm^2\tau^2}\sum_{i=0}^{M-1}
  \Biggl[\sum_{k=0}^{m-1}\Biggr.\nonumber\\
  &\qquad
  \Biggl.\Bigl(\frac{m-1}{2}-k\Bigr) 
  \Bigl(\mathsf{x}_{i+k}-\mathsf{x}_{i+m+k}\Bigr)\Biggr]^2~,
\end{align}
with $M=N-2m$, since $w(t)$ spans over $2m$ samples.
\Enrico{The main advantage of PVAR is that the weight applied to the $\mathsf{x}_j$ samples is equivalent to a linear regression, which features the least-squares fit of the slope.  PVAR$(\tau)$ is therefore an estimator of the variance of the slope of the $\mathsf{x}_j$ samples over the duration $\tau$. For a detailed description of PVAR and its properties see the original article \cite{Vernotte-2016-TUFFC}}

\subsection{Response of AVAR and PVAR to $f^\alpha$}
The response of a generic $\sigma^2_\mathsf{y}(\tau)$ to $S_\mathsf{y}(f)$ is
\begin{align}
  \sigma^2_\mathsf{y}(\tau) = 
  \int_0^\infty \left|H(f)\right|^2 S_\mathsf{y}(f)\:\mathrm{d}f\:,
  \label{eqn:resp_int}
\end{align}
where $H(f)$ is the transfer function, or
\begin{equation}
  \sigma^2_\mathsf{y}(\tau,\alpha) = 
  \int_0^\infty \left|H(f)\right|^2 h_\alpha f^\alpha\:\mathrm{d}f
  \label{eq:resp_int}
\end{equation} 
for the $\alpha$-th term of the polynomial law \eqref{eqn:poly-law}.
Using the subscript $A$ for AVAR, $\left|H(f)\right|^2$ becomes 
\begin{align}
  \label{eqn:HA}
  \left|H_A(f)\right|^2 
  &=\frac{2\sin^2(2\pi f\tau)}{\left(\pi f\tau\right)^2}~,
\end{align}
therefore
\begin{align}
  \label{eqn:avar_response}
  \mathrm{AVAR}(\tau,\alpha)
  &=\frac{\left(2^{-\alpha+1}-4\right)\Gamma(\alpha-1)\sin(\pi\alpha/2)}{(2\pi \tau)^{\alpha+1}}\:\mathsf{h}_\alpha\:.
\end{align}
This is equivalent to \cite[Eq.~(14)]{Walter-1994} because we have not introduced the usual cutoff frequency $f_H$ in \eqref{eqn:resp_int}. 

Similarly, the transfer function associated to PVAR is
\begin{align}
  \label{eqn:HP}
  \left|H_P(f)\right|^2
  &= \frac{9\left[2\sin^2(\pi f\tau)-\pi\tau f\sin(2\pi f\tau)\right]}{2\left(\pi f\tau\right)^6}~,
\end{align}
which is Eq. (17) of \cite{vernotte2016}. Combining Eq. (\ref{eq:resp_int}) and (\ref{eqn:HP}), we derive the response of PVAR
\begin{align}
  \mathrm{PVAR}(\tau,\alpha) 
  &=9{\times}2^{5-\alpha}
  \Bigl[\alpha^2-\alpha -4 -2^\alpha(\alpha-3)\Bigr]\times{}\nonumber\\
  &\qquad\times\frac{\Gamma(\alpha-5)\:\sin(\pi\alpha/2)}{(2\pi\tau)^{\alpha+1}}\:\mathsf{h}_\alpha\:.
  \label{eqn:pvar_response}
\end{align}
Because PVAR converges for $f^\alpha$ from $f^{-2}$ to $f^{+2}$ FM noise, we can assume that \eqref{eqn:pvar_response} is valid for $\alpha\in{}]{-3,+3}[$.

\begin{figure}
\includegraphics[width=\linewidth]{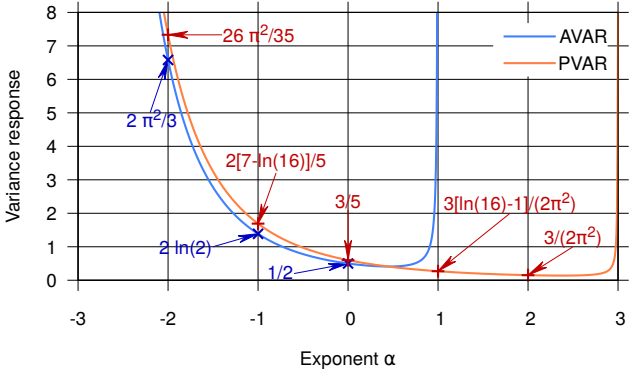}
\caption{Continuous response of AVAR and PVAR compared to the known responses for $\alpha\in{}]{-3,+3}[$. The responses of AVAR are not plotted for $\alpha\geq 1$ because this estimator diverges without the introduction of a high cut-off frequency.}
\label{fig:reponses}
\end{figure}

Figure~\ref{fig:reponses} shows AVAR and PVAR calculated from Eq. (\ref{eqn:avar_response}) and (\ref{eqn:pvar_response}), as a function of $\alpha$.  For integer $\alpha$, the results are the same as in \cite[Table I]{vernotte2016}.

\section{Degrees of Freedom of PVAR estimates}
First, we have to find a simple expression for the number of degrees of freedom (dof) of PVAR estimates for integer power-law noises. 
\cor{ Since an equation has been found for a white PM noise (see Eq. (24) in \cite{vernotte2016}), we assume that it can be generalized for other noise types 
 to}
\begin{equation}
  \nu\approx\frac{35}{A(\alpha)m/M-B(\alpha)(m/M)^2}
  \label{eq:approx}
\end{equation}
where $A(\alpha)$ and $B(\alpha)$ are coefficients that need to be determined. From \cite[Eq. (24)]{vernotte2016}, we already know that $A(+2)=23$ and $B(+2)=-12$.
We determine the general $A(\alpha)$ and $B(\alpha)$ from massive Monte-Carlo simulations, and verify the results by comparing them to the dof computed for a continuous power-law.

\subsection{Determination of the Coefficients from Monte-Carlo Simulations}

\begin{figure}[b!]
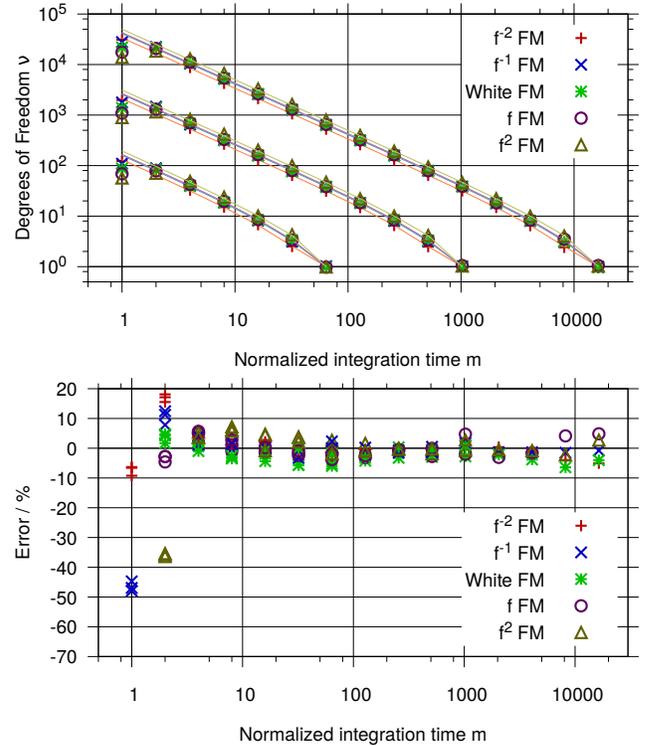

\includegraphics[width=\linewidth]{all_VoV.pdf}
\includegraphics[width=\linewidth]{all_VoV_err.pdf}
\caption{Above: comparison of the empirical dof (crosses) and the approximations (lines) given by Eq.~\eqref{eq:approx} and \eqref{eq:coeff_A} for all types of noise. \cor{The lines are drawn from point to point without interpolation and the last point is set to 1.} 
Below: relative difference (in \%) between the empirical dof and the approximations. \label{fig:dof}}
\end{figure}

The Monte-Carlo simulation was performed by computing 10\,000 sequences of frequency deviations for each $\alpha \in \left\{-2, -1, 0 , +1, +2\right\}$ and for different data length $N \in \left\{128, 2048, 32768\right\}$, i.e. 150\,000 simulated sequences in total. For a given $\alpha$, $N$ and $\tau$, we derived the dof from the averages and the variances of the PVAR for the corresponding set of sequences by using the following well-known property of $\chi^2_\nu$ distributions \cite{vernotte2016}: 
\begin{equation}
	\nu=2\frac{\mathbb{E}^2[\textrm{PVAR}(\tau)]}{\mathbb{V}[\textrm{PVAR}(\tau)]},
	\label{eq:var2dof}
\end{equation}
where $\mathbb{E}[~]$ and $\mathbb{V}[~]$ are the mathematical expectation and the variance of the argument.
The least square fit results in
\begin{itemize}
	\item $A(-2)\approx 34$, $A(-1)\approx 28$, $A(0)\approx 27$, $A(+1)\approx 27$, $A(+2)=23$
	\item $B(\alpha) \approx 12$ for all $\alpha$.
\end{itemize}
We have then modeled $A(\alpha)$ by the following $3^\textrm{\footnotesize rd}$ order polynomial and assumed that $B(\alpha)=B$ is constant:
\begin{equation}
  \begin{array}{l}
  A(\alpha)=27+\frac{1}{4}\alpha+\frac{5}{14}\alpha^2-\frac{3}{4}\alpha^3\\[1ex]
B=12.
\end{array}\label{eq:coeff_A}
\end{equation}
Using Eq.~\eqref{eq:approx} and \eqref{eq:coeff_A}, we are now able to assess the dof of all PVAR estimates regardless of the normalized integration time $m$ or the number of samples $M$.

The upper plot of Fig. \ref{fig:dof} compares the dof obtained by the Monte-Carlo simulations and by Eq.~\eqref{eq:approx} and \eqref{eq:coeff_A} for all integer types of noise. The agreement is confirmed by the lower plot which shows that the discrepancies are within $\pm10 \%$ except for the very first values of $m$ ($m=1,2$).

The model provided by \eqref{eq:approx} and \eqref{eq:coeff_A} can be applied to the classical power law, with integer $\alpha$.  Next, we check 
 its validity as an extension for real $\alpha \in ]-3,3[$ by computing the dof of PVAR.

\subsection{Verification for Continuous Polynomial-Law Noise}
The dof can be computed from Eq.~\eqref{eq:var2dof}. The mathematical expectation of the response of PVAR is given by \eqref{eqn:pvar_response}, and the variance can be computed from (21) and (22) of \cite{vernotte2016}
\begin{multline}
	\mathbb{V}\left[ \textrm{PVAR}(\tau )\right] = \frac{2}{M^{2}}\sum_{i=0}^{M-1}\sum_{j=0}^{M-1}\left[\frac{72}{m^4\tau^2}\right.\\
	\sum_{k=0}^{m-1}\sum_{l=0}^{m-1}\left(\frac{m-1}{2}-k\right)\left(\frac{m-1}{2}-l\right)\\
	\Bigl\{2R_\mathsf{x}[(i+k-j-l)\tau_{0}]\Bigr.\\
	-R_\mathsf{x}[(i+k-j-m-l)\tau_0]\\
	\left.\Bigl.-R_\mathsf{x}[(i+m+k-j-l)\tau_{0}]\Bigr\}\vphantom{\frac{72}{m^4\tau^2}}\right]^2
	\label{eq:var_of_pvar}
\end{multline}	
where $R_\mathsf{x}(\tau)$ is the autocorrelation function of the phase-time 
 $\mathsf{x}(t)$, i.e., $R_\mathsf{x}(\tau)=\xp{\mathsf{x}(t)\mathsf{x}(t+\tau)}$. We use the following continuous expression of $R_\mathsf{x}(\tau)$ versus the power-law exponent $\alpha$ (see \cite{Kasdin-1995,Walter-1994}):
\begin{equation}
R_\mathsf{x}(m\tau_0)=\frac{\mathsf{h}_\alpha}{2(2\pi)^\alpha\tau_0^{\alpha-1}} \frac{\Gamma(m-\alpha/2+1)\Gamma(\alpha-1)}{\Gamma(m+\alpha/2)\Gamma(\alpha/2)\Gamma(1-\alpha/2)}.\label{eq:gamma_m}
\end{equation}
\Enrico{Because this expression involves the $\Gamma$ function with argument of the order of $m$, the computations is practically limited to $N=128$ samples (notice that $\Gamma(128)\simeq3{\times}10^{213}$). 
This difficulty is avoided using the property that $\Gamma(z)=(z-1)\Gamma(z-1)$ for $z>1$, and the recursive formula}
\cor{%
$$
\frac{\Gamma(m-\alpha/2+1)}{\Gamma(m+\alpha/2)}=
\frac{\Gamma(3-\alpha/2)}{\Gamma(2+\alpha/2)}\prod_{j=0}^{m-3}\frac{m-j-\alpha/2}{m-1-j+\alpha/2}
$$}%
\Enrico{where the arguments of $\Gamma$ are greater or equal to 1 for $\alpha \in [-2,+2]$.}
\cor{Therefore, the autocorrelation function can be computed for large $N$ as
\begin{eqnarray}
R_\mathsf{x}(m\tau_0)&=&\frac{\mathsf{h}_\alpha}{2(2\pi)^\alpha\tau_0^{\alpha-1}} \frac{\Gamma(3-\alpha/2+1)\Gamma(\alpha-1)}{\Gamma(2+\alpha/2)\Gamma(\alpha/2)\Gamma(1-\alpha/2)}\nonumber\\
&&\times \prod_{j=0}^{m-3}\frac{m-j-\alpha/2}{m-1-j+\alpha/2}.
\end{eqnarray}}%
\Enrico{We used this equation to compute the theoretical variance of PVAR$(\tau)$ versus the continuous variable $\alpha$, and we deduced the dof from \eqref{eq:var2dof}}.

\begin{figure}
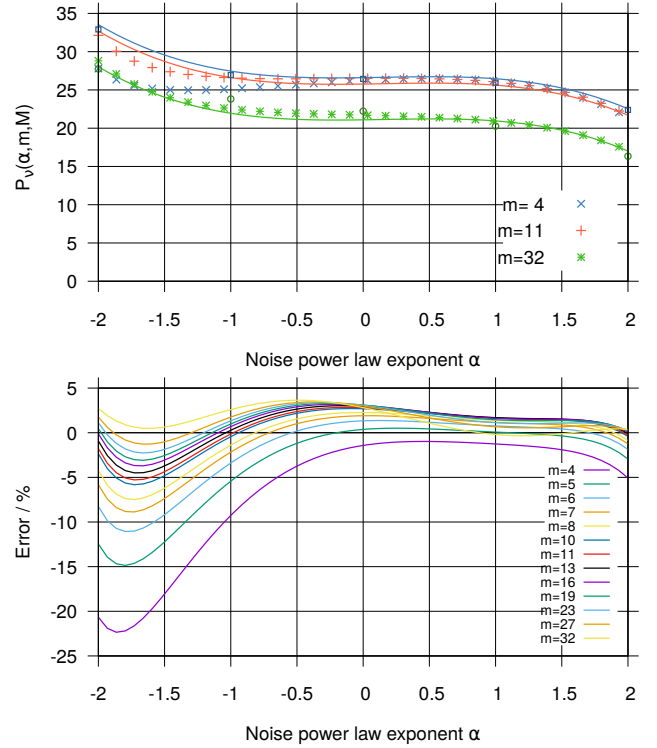

  \includegraphics[width=\linewidth]{contnoise_dof_fit.pdf}
  \includegraphics[width=\linewidth]{contnoise_dof_fit_err_alltau.pdf}
	\caption{Above: comparison of $P_\nu(\alpha,m,M)$ computed from (\ref{eq:var_of_pvar}) ($\times,+,*$) and approximated by (\ref{eq:approx}) (solid lines) for $N=128$ data. The blue squares and the green circles are respectively the values obtained for $m=4$ and $m=32$ from the Monte-Carlo simulations.  Below: Error (in \%) between the approximated values of $P_\nu(\alpha,m,M)$ and the computed values. 
  \label{fig:dof_vs_alpha}}
\end{figure}

Let us define $P_\nu(\alpha,m,M)=\frac{35 M}{m\nu}$.  From Eq.~\eqref{eq:approx}, we see that $P_\nu(\alpha,m,M)\approx A(\alpha)-Bm/M$. The top plot of Fig.~\ref{fig:dof_vs_alpha} shows $P_\nu(\alpha,m,M)$ computed from \eqref{eq:var_of_pvar} (crosses) and approximated from \eqref{eq:coeff_A} (solid lines) versus the noise power-law $\alpha$ for $m\in\{4,11,32\}$ (we prefer to plot $P_\nu(\alpha,m,M)$ instead of $\nu$ for a better visualisation).  The agreement is quite good for $m=11$ and $m=16$, but there is a notable difference for $m=4$ and $\alpha <-1$. The lower plot of Fig.~\ref{fig:dof_vs_alpha} shows that this discrepancy is of $\approx20\%$ maximum, but it remains within $\pm5$\% in most cases (all $\alpha$ for $m>8$, and all $m$ for $\alpha>-1$). This agreement is satisfactory to get an acceptable assessment of the PVAR uncertainties since the relative uncertainties are proportional to $1/\sqrt{\nu}$: they are therefore always below 10\% and mostly within $\pm2.5\%$. 

\subsection{The Case of the Largest Integration \cor{T}ime}
\begin{figure}
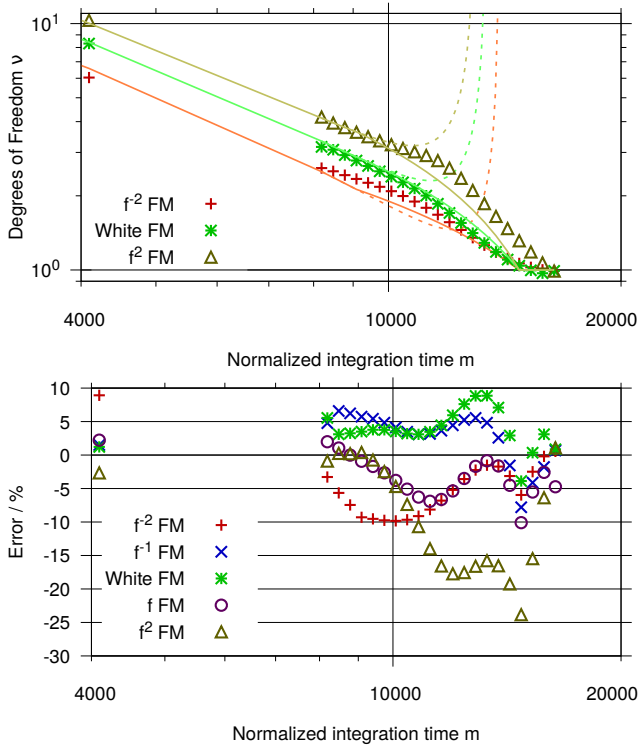

\includegraphics[width=\linewidth]{all_VoV_end_32768.pdf}
\includegraphics[width=\linewidth]{all_VoV_err_end_32768.pdf}
\caption{Above: comparison of the empirical dof (crosses) and the semi-logarithmic fits (solid lines) for $N/4 \leq m \leq N/2$ and for random walk FM, white FM and white PM. The dashed lines represent the approximations given by Eq. (\ref{eq:approx}) and (\ref{eq:coeff_A}). In this example $N=32768$ samples and the logarithmic increment of the $m$-values is $2^{1/20}$ within $[N/4,N/2]$.
Below: Error (in \%) between the empirical dof and the semi-logarithmic fits for all types of noise. \label{fig:dof_end_vs_alpha}}
\end{figure}

The approximation given by Eq. (\ref{eq:approx}) and (\ref{eq:coeff_A}) is close enough to the empirical dof $\nu$ for $m\leq N/4$. Moreover, we know that $\nu=1$ for $m=N/2$.  \cor{This is enough to draw Fig. \ref{fig:dof} 
 since no interpolation is performed between the last 2 points, i.e. $m=N/4$ and $m=N/2$.} On the other hand, we note that the approximation diverges beyond $N/4$ (dashed lines in the upper plot of Fig.~\ref{fig:dof_end_vs_alpha}) \Enrico{, if intermediate values of $m$ are computed}. However, it is important to assess the uncertaintiy within this interval, particularly if $N$ is not a power of 2. 

We fill this gap by interpolating the dof within $\textrm{round}\left(2^{3/20}N/4\right) \leq m \leq \textrm{round}\left(2^{-3/20}N/2\right)$ (rounding is necessary to ensure that $m$ is an integer), i.e. between $m_1\approx\textrm{round}(1.11 N/4)$ and $m_2\approx\textrm{round}(0.901 N/2)$, with the following semi-logarithmic fit
\begin{equation}
	\nu(m)=a \ln(m) +b 
\end{equation}
with
\begin{align}
	a&=\frac{\nu(m_1)-1}{\ln(m_1)-\ln(m_2)}\\[1ex]
  b&=\frac{\ln(m_1)-\nu(m_1)\ln(m_2)}{\ln(m_1)-\ln(m_2)}.
\end{align}
For $m\geq m_2$, the dof are set to 1.

\cor{ To focus on the result of the semi-logarithmic fit an enlargement of the highest 2 decades of $m$, i.e. $m\in[4096,8192]$ for $N=32768$ data is shwon in the top of Fig.~\ref{fig:dof_end_vs_alpha}.} The bottom plot shows the error between the fit and the dof computed from the Monte-Carlo simulations. Most of these errors are within $\pm10\%$, except for  white FM\@.  In this case of white FM, the error is between $+5\%$ and $-20\%$, and up to $-24\%$ for $m = 14\,766$. However, this fit is sufficient to ensure an estimation of the PVAR uncertainty for the highest $\tau$ within $\sim10\%$ at worst.

\section{Conclusion}
We have determined the response of PVAR for continuous power-law noise spectra from a theoretical calculation.  Using Monte-Carlo simulations, we have obtained a simplified expression providing the dof of the PVAR estimates within 10~\%. We have proven that this expression remains valid for non-integer power-law noise types.  Finally, we have shown that a simple interpolation is efficient to fit the dof for the highest octave of integration times. \Enrico{These results generalize the use of the PVAR to process signals with a non-integer powers in the polynomial-law spectrum}. \cor{This can be used to analyze the timing of milliseconds pulsars and to estimate the non-integer exponent of a red noise, if it is detected.}

\section*{Acknowledgements}
This work is funded by the ANR \selectlanguage{french}Programme d'Investisse\-ment d'Avenir\selectlanguage{english} (PIA) under the FIRST-TF network (ANR-10-LABX-48-01), the Oscillator IMP project (ANR-11-EQPX-0033-OSC-IMP) and the EUR EIPHI Graduate School (ANR-17-EURE-00002), and by grants from the Région Bourgogne Franche Comté intended to support the PIA.

\bibliographystyle{IEEEtran}
\bibliography{Ref-short,Ref-Rubiola,References,Ref-local}


\end{document}